# MODELLING HIGH RESOLUTION ECHELLE SPECTROGRAPHS FOR CALIBRATIONS

**Hanle Echelle spectrograph, a case study**

**Anantha Chanumolu [a], Damien Jones [b], Sivarani Thirupathi [a]**

[a] *Indian Institute of Astrophysics, 2nd block Koramangala Bangalore 560034, India*
*E-mail*: anantha@iiap.res.in, sivarani@iiap.res.in

[b] *Prime Optics, Eumundi Q 4562 Australia*
*E-mail:* damien@primeoptics.com.au

ABSTRACT: We present a modelling scheme that predicts the centroids of spectral line features for a high resolution Echelle spectrograph to a high accuracy. Towards this, a computing scheme is used, whereby any astronomical spectrograph can be modelled and controlled without recourse to a ray tracing program. The computations are based on paraxial ray trace and exact corrections added for certain surface types and Buchdahl aberration coefficients for complex modules. The resultant chain of paraxial ray traces and corrections for all relevant components is used to calculate the location of any spectral line on the detector under all normal operating conditions with a high degree of certainty. This will allow a semi-autonomous control using simple in-house, programming modules. The scheme is simple enough to be implemented even in a spreadsheet or in any scripting language. Such a model along with an optimization routine can represent the real time behaviour of the instrument. We present here a case study for Hanle Echelle Spectrograph. We show that our results match well with a popular commercial ray tracing software. The model is further optimized using Thorium Argon calibration lamp exposures taken during the preliminary alignment of the instrument. The model predictions matched the calibration frames at a level of 0.08 pixel. Monte Carlo simulations were performed to show the photon noise effect on the model predictions.

# 1. Introduction

In the recent years there has been considerable improvement in the designing and building of spectrographs. White pupil configurations (Baranne et al. 1996) have improved the system efficiencies. Use of optical fibers for feeding spectrographs has allowed instrument to be placed in environment stable enclosures away from the telescope (Pepe et al. 2000). Image slicers (Raskin et al. 2011) and anamorphic pupil slicers (Spano et al. 2012) improved the efficiencies for narrow slit widths. Precision thermal and pressure controlled environment of spectrographs, vacuum enclosures improved the stability up to 1m/s (Pepe et al. 2000). Laser frequency combs are being developed for better wavelength calibrations (Wilken et al. 2012), improving the radial velocity accuracies to few tens of centimeters. However, little has changed in the way data is calibrated. Once the instrument is built and installed, the physics that went into the designing and building of the instrument is seldom put to use in its calibrations, scientific operations and quality control of the instrument (Bristow et al. 2008). The ESO group has successfully implemented a model based calibration approach to instruments such as FOS and STIS on HST, CRIRES and X-shooter on VLT (Rosa et al. 2002; Kerber et al. 2005; Bristow et al. 2006; Bristow et al. 2008). The earlier work by the ESO team uses a modelling approach that is on-axis limit with the all aberration corrections applied together as a polynomial at the detector plane (Bristow et al. 2006). The scheme we present here, aims at providing the exact position of chief rays on the detector plane leaving only the centroid corrections, which are of the order of sub-pixel as a correction at the detector plane. This reduces error propagation through the system, and the degeneracy states for the optimization routine to predict the instrument status. This is achieved using paraxial ray tracing and including exact corrections for certain surface types and Buchdahl aberration coefficients for complex modules such as camera systems.

Thorium Argon calibration images will be used along with a simulated annealing optimization routine to derive the model parameters that match with the built instrument. The optimized model can be used for understanding instrument behaviour, quality checks, and trending. Apart from this, we expect the model will help in accurate wavelength calibrations, 1D extraction of orders and the shift correction for simultaneous ThAr observations for radial velocities. The model once matched with the instrument can predict the location of any line on the detector given the position on slit and the wavelength, and so can be used for wavelength calibrations, this will help in the low density ThAr emission line regions. The tilt of the slit image can be well predicted by the model, which improves the accuracy of 1D extraction especially long sliced slit. Radial velocity observations with simultaneous ThAr assumes a constant shift throughout the spectrum and that the star and calibration light follow the same path. A model will be able to predict the position dependent shift in the spectrum and also the instrument dependent shift in star spectrum using the calibration spectrum.

The modelling effort for Hanle Echelle SPectograh (HESP) was done in parallel with the development, fabrication and testing of the instrument. This gives an advantage of better knowledge of various components of the instrument that need to be included in the model, with minimum assumptions. Measured values of the fabricated components, melt data of the glasses and opto-mechanical surface measurements were used in the model. These measured values help to estimate the bounds for model parameters that are to be optimized.

In the paper, section-1 describes the Hanle Echelle spectrograph and its components and in section- 2 formalism of the model is presented. The necessary mathematics in building each subsystem of the model is presented. The simplification used in modelling the camera is also presented in detail. In Section 3, we present the comparison of the model with the commercial ray trace software, ZEMAX. We also describe how various corrections were derived in a modular way to match the model to the real instrument. Finally Section 4 presents the preliminary results of the performance of the model with the real instrument followed by a summary and conclusion. The improvement of model based calibration over empirical calibration of precise RV and accurate extraction will be presented in a future paper.

# 2. Hanle Echelle Spectrograph (HESP)

Hanle Echelle Spectrograph (HESP) is being developed as collaboration between Kiwistar optics, Callaghan Innovation, New Zealand and Indian Institute of Astrophysics, Bengaluru, India. This project is supported by DST grant IR/S2/PF~02/2010 under IRHPA scheme. The spectrograph is expected to be commissioned in July 2015 at the 2m Himalayan Chandra telescope, Hanle. The spectrograph is designed to operate in two resolutions, R=30000 and 60000, in white pupil configuration for increased efficiency. The spectrograph is fed through a 27m optical fiber link from the cassegrain port at the telescope to the bench mounted spectrograph that is located inside a thermal enclosure in the ground floor of the observatory.

Specifications of the spectrograph are:
Wavelength Coverage: 350nm to 1000nm on single detector over 64 orders
Resolution: 30000 and 60000
Two Fibres: Star-sky observations and provision for simultaneous reference
observations using ThAr lamp
Non-referenced mechanical stability: 200m/s
Simultaneous- reference Velocity precision: 20 m/s

### 2.1 Cassegrain Unit

Fig. 1 shows the optical and mechanical layout of the Cassegrain unit of the system. Light from the telescope passes through the atmospheric dispersion correctors (ADCs) before coming to focus at the pinhole mirror. The ADCs are two counter rotating prism doublets to correct for the atmospheric dispersion at different zenith angles ranging from $0^0$ to $70^0$. At the focal plane, a pinhole mirror is placed that has two pinholes to allow light from the star and the sky. The spilled over starlight, off the 2.85" pinhole will be directed to an acquisition and guiding unit (A&G unit). The F/9.2 telescope beam is converted to F/3.6 beam before being fed into the optical fibres. The calibration fibres, from the calibration unit of the Spectrograph room feed the light into the spectrograph fibres through a beam splitter placed in the beam path as shown in the Fig. 1.

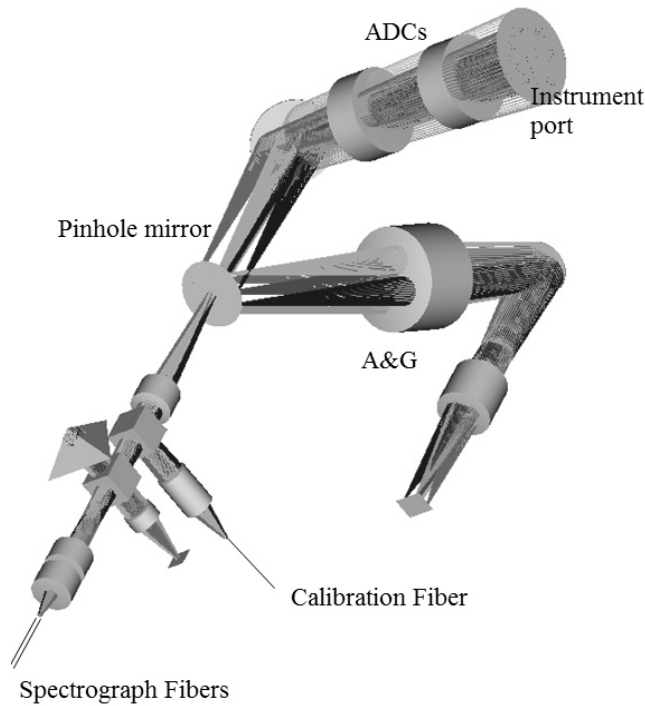

**Fig. 1** Cassegrain unit Optical layout

### 2.2 Spectrograph Unit

The optical fibres end at the input optics module of the spectrograph. Fig. 2 shows the optical layout of the input optics and the spectrograph. Beam from the fibres is converted to F/75. The optical fibres along with these conversion optics are mounted on a linear translation stage to switch between R=30000 and 60000 resolution modes. In the case of R= 60000 resolution mode, the F/75 beam passes through an image slicer that slices the 2.85 arc sec fiber image into two 1.25 arc sec slices. The beam is then converted into F/10.45 which focuses at the slit plane of the spectrograph.

A 700mm diameter parabolic mirror collimates the light from the input optics before the beam falls on the Echelle grating. The Echelle grooves are horizontal, dispersing the beam in vertical direction. The collimator mirror in second pass forms an intermediate image on the slit mirror. In the third pass the collimator collimates the beam again. The fold mirror directs the beam into the cross disperser prisms. The first prism is placed at the white pupil of the system. Two prisms with apex angles of 57 degrees disperse the beam in the horizontal direction separating the different Echelle orders. A five element camera system, images the dispersed light onto the detector. The final element of the camera optics is a field flattener, which along with the detector is placed at about 5 degrees angle to avoid possible ghost images.

The Detector is a 4K x 4K E2V CCD231-84, with 15 micron pixel size. The detector has a custom graded AR-coating, that matches the wavelength format of the spectrograph, in order to increase the efficiency. The average FWHM of the slit image on the detector is 4pixels, in the low resolution mode, whereas it is 2pixels in the high resolution mode.

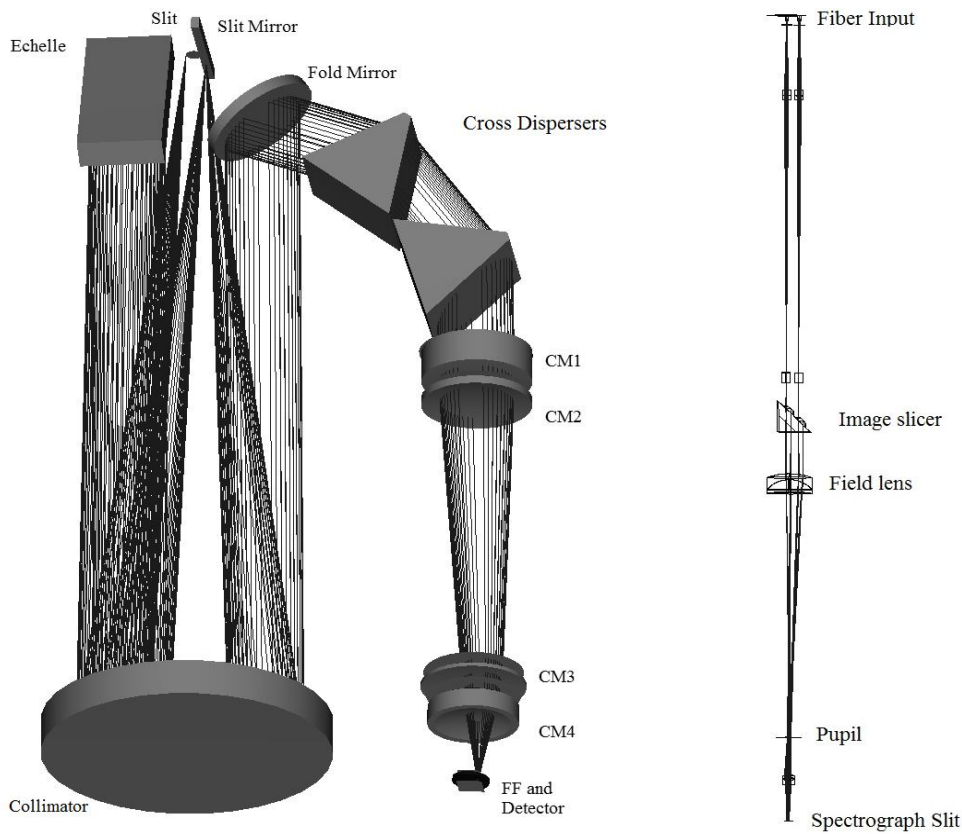

**Fig. 2** Optical layout of the Hanle Echelle Spectrograph

## 3. Physical Model

In this section we present the methodology followed to model the spectrograph. An Object oriented approach is used to model the system making it possible to adapt this implementation for other spectrographs with a few changes. Every relevant component of the system is described in a class with the component based parameters described as its attributes and the transformation functions as the methods of the class. Using the defined component classes and ray traces between them, the final instrument will be modelled which will be described in detail later. Fig. 3 shows an example of the class diagram of a prism.

Different Modules relevant to HESP are:

1. Slit
2. Collimator
3. Echelle Grating
4. Prism
5. Camera
6. Detector

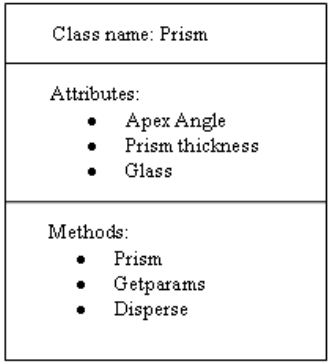

Fig. 3 Class Diagram of Prism

### 3.1 Coordinate System and Ray coordinates

Fig. 4 shows the coordinate system used for the model. Looking from the positive axes, clockwise direction is defined positive. The different components' have their optic axis along the z –axis. A Ray (OP) is defied by its coordinates (x, y, z) at the point of intersection on a surface and cosine of angles ($\alpha,\beta,\gamma$) it makes with the x, y and z axes, (DCx, DCy, DCz) respectively and the wavelength of the ray, $\lambda$.

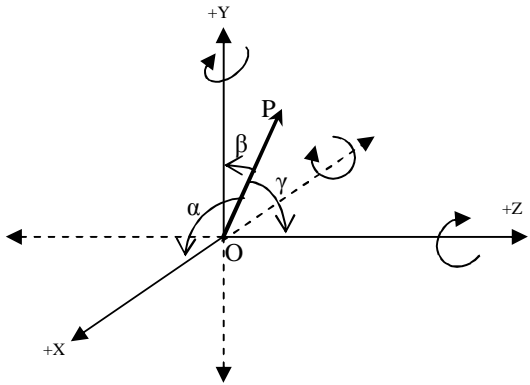

Fig. 4 Coordinate System Definition

### 3.2 Class Definitions

1. Slit: The slit is described by its position (x, y), the angles of orientation ($\mu_x$, $\mu_y$, $\mu_z$), slit width and height.
2. Collimator: The mirror collimator is defined by its curvature and the conic constant. The inputs to the transformation function are the ray coordinates at the paraxial plane of the collimator, P (Fig. 5). The curvature *(c)* and conic constant *(k)* are used to establish the sag equation of the mirror surface. The coordinates at A (Fig. 5) on the mirror surface are calculated using paraxial surface coordinates using the following:

Sag equation: $$dz = \frac{(c_x x^2 + c_y y^2)}{1+\sqrt{1-(1+k_x)c_x^2 x^2 - (1+k_y)c_y^2 y^2}}$$

*($c_x, c_y$) are the curvatures in x and y*
*($k_x, k_y$) are the conic constants in x and y*

For a parabolic mirror dz can be calculated by

$$d\eta = \frac{-(x_p dT_x + y_p dT_y) - \frac{r1}{2} - \left(\frac{r1}{|r1|}\sqrt{\left(\frac{r1}{2}\right)^2 - r1 x_p dT_x - r1 y_p dT_y - (y_p dT_x - x_p dT_y)^2}\right)}{dT_x^2 + dT_y^2}$$

Where
$(x_p, y_p)$ → ray intersect coordinates on the paraxial collimator plane
$(dT_x, dT_y)$ → Direction tangents of the ray
$r1$ → twice the radius of curvature of sag
$dz$ → z distance from the paraxial plane to the actual surface

The coordinates of the ray intersection $(x_{new}, y_{new})$ on the mirror surface are given by
$$x_{new} = x_p - (d\eta \times dT_x)$$
$$y_{new} = y_p - (d\eta \times dT_y)$$

The direction cosines of the reflected ray at this point will be
$$dc_{new} = dc_{old} - (2\cos\theta) normal$$

Where
$dc_{new}, dc_{old}$ are the direction cosines of the ray after and before reflection
$\theta$ is the angle between the ray and normal at the point $(x_{new}, y_{new})$.

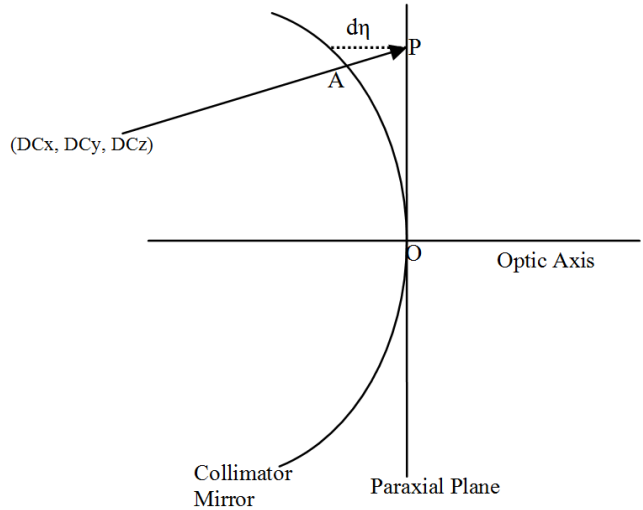

**Fig. 5** Two Dimensional Depiction of Collimator Mirror

3. Grating: Grating is defined by its grating constant and direction of grooves. The transformation function to calculate the diffracted rays' direction vector uses the prescription by Mitchell 1981.

4. Prism: The attributes to a prism are its apex angle (α), dispersion equation of the glass (n(λ,T)) and its base thickness (d). The transformation function is the 3D refraction at the surfaces and ray trace within the prism. Inputs to the prism transformation function are the (x,y) coordinates of the ray intercept on the plane AB shown in the Fig. 6 and the wavelength of the ray.

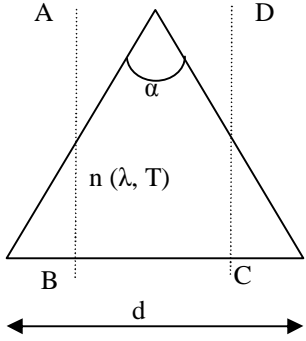

**Fig. 6** Prism Description

The output of the transformation function is the coordinates of the ray on CD in Fig. 6. The 3D refraction at a surface is given by

$$dc_{new} = \frac{n1}{n2} dc_{old} + \left(\cos r - \frac{n1}{n2} \cos i\right) normal$$

Where
$dc_{new}$, $dc_{old}$ → direction vector of the ray after and before refraction
$n1$, $n2$ → refractive indices of medium before and after the refractive surface
$i$, $r$ → Angle made by incident ray and refracted ray with normal to the surface
at the point of intersection of ray with the surface

5. Camera: Camera, in general a multi-element system, is designed to correct the various aberrations introduced by the spectrograph optics. HESP camera is a five element system, CM1 to FF in Fig. 2. The first four elements are rotationally symmetric whereas the fifth element is a field flattener with one cylindrical face and placed at an angle with respect to the first four elements to eliminate the ghost imaging.

    A regular ray trace from surface to surface in a camera system will need a number of square roots and trigonometric calculations. Also a model with a reasonable trade off between number of parameters to optimize when matching with the built instrument and accuracy is needed. In order to facilitate the same following methodology is implemented. The first four elements are modelled as a single unit. The unit is described by its ABCD matrix and aberration polynomial corrections to the output. The output from this unit is traced through the single element field flattener for the best accuracy at the detector plane.

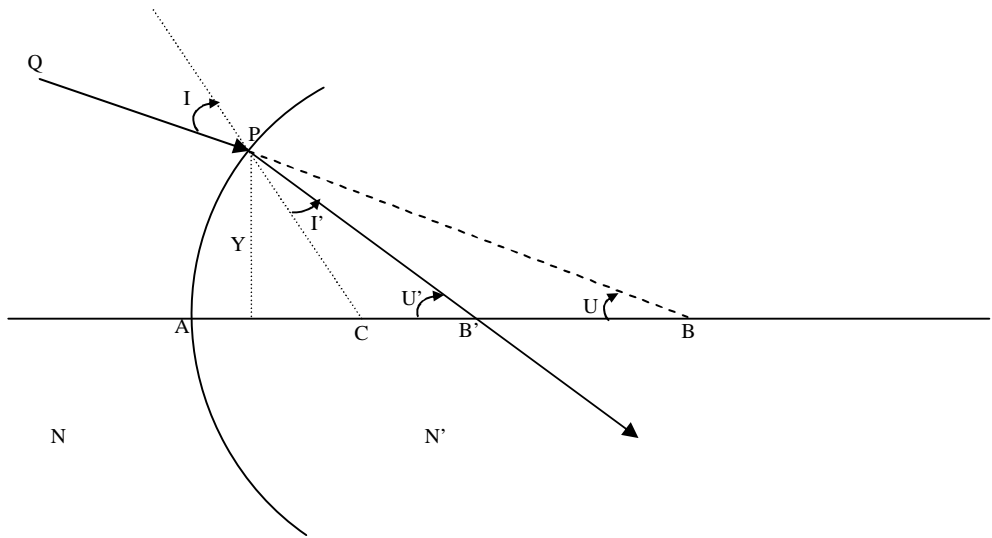

**Fig. 7** Refraction at a spherical surface

*ABCD Matrices: Paraxial coefficients of Lens system*
In reference to Cruickshank et al. 1960, Fig. 7 shows refraction of a ray QP at a spherical surface. The ray QP falling on spherical surface with curvature, *c,* with vertex at A and optical axis AC, at angle I to the normal CP is refracted as PB'. The incident ray makes an angle U and the refracted ray makes an angle U' with the optical axis. The incident ray hits the spherical surface at a height Y from the optical axis. The refractive indices of the medium before and after the refracting surface are N and N' respectively. In paraxial approximation,

$$\mu' = \mu + \phi Y$$
$$\mu = NU, \mu' = N'U'$$
$$where, \phi = (N'-N)c,$$

The coordinates *(Y,μ)* will be used to describe a ray at the surface and (Y, *μ'*) are the coordinates of the refracted ray. To trace a ray through a series of refracting surfaces, the ray will be described at the first surface by *(Y₁,μ₁)*, at surface 2 by *(Y₂, μ₂)* and so on.
Also,

$$\mu_2 = \mu_1'$$
$$Y_2 = Y_1 - \tau_1' \mu_1'$$

where, $\tau_1'$ is the distance between first and second surfaces.
At any given surface *k*, the coordinates can be calculated by,

$$Y_k = A_k Y_1 + B_k \mu_1$$
$$\mu_k' = C_k Y_1 + D_k \mu_1$$

Where $A_k$, $B_k$, $C_k$ and $D_k$ are constants determined by radii, thickness and refractive indices. The coefficients can be calculated iteratively by,

$$A_{j+1} = A_j - \tau_j' C_j$$
$$B_{j+1} = B_j - \tau_j' D_j$$
$$C_{j+1} = C_j + \phi_{j+1} A_{j+1}$$
$$D_{j+1} = D_j + \phi_{j+i} B_{j+1}$$
$$A_1 = 1, B_1 = 0, C_1 = \phi_1, D_1 = 1$$

So the paraxial coordinates of a ray at the Gaussian image plane for a system can be calculated just by the incident ray coordinates without the intermediate calculations every time.

*Aberration Polynomials*
Using ABCD matrices the Gaussian image position can be calculated. But a real optical system doesn't perform ideal imaging. All rays emerging from a single object point do not converge to single image point. The difference between the ideal image position and the real ray's intersection point on the Gaussian image plane, called the transverse aberrations, need to be established for the camera system to know the ray's output coordinates to a high accuracy.

Fig. 8 is paraxial representation of an axially symmetric system. F1 is the object plane of the optical system whose first and last surfaces' polar tangent surfaces are shown in the Figure. One of the rays from object point $O_1$ pointed at P on the entrance pupil is also depicted in the figure. Within paraxial regime, all the rays emerging from object point $O_1$ with coordinates $(x_1, y_1, l_{o1})$ in the object plane $F_1$ converge at the ideal image plane $F'_k$ at the point $O'_k$ with coordinates (x'$_k$, y'$_k$, l'$_{ok}$). If a real ray strikes the plane F'$_k$ at $O_k$ with coordinates $(x_k, y_k, l'_{ok})$, then transverse aberration is given by, (Fig. 9)

$$\varepsilon_{xk} = x'_k - x_k$$
$$\varepsilon_{yk} = y'_k - y_k$$

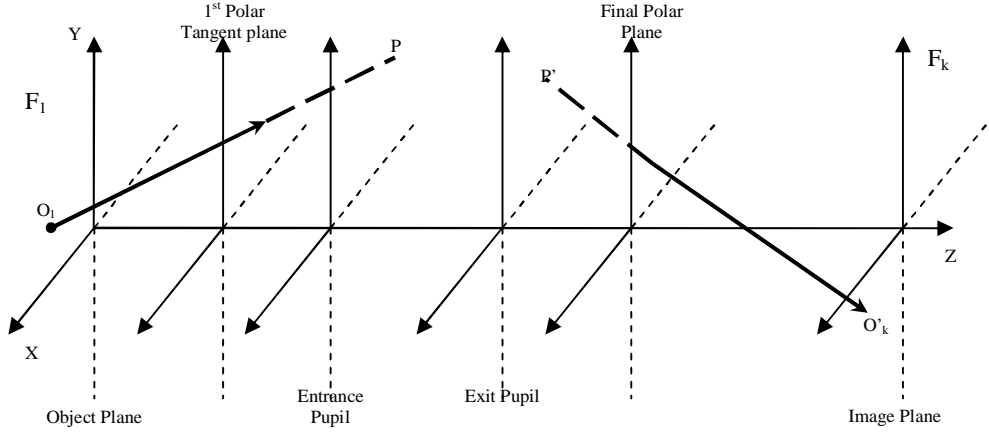

**Fig. 8** Paraxial Representation of a rotationally symmetric camera system

($\varepsilon_{xk}$, $\varepsilon_{yk}$) can be expanded in a series of ascending powers of h (radial distance of object point $O_1$) and polar coordinates of the paraxial ray on the entrance pupil P(R, θ) shown in Fig. 9 . The series is a sum of set of homogeneous polynomials of degree 3, 5, 7, …, (2n+1), … . The coefficients of the terms are referred to as third-order, fifth-order, seventh-order, etc aberration coefficients (Cruickshank 1960).

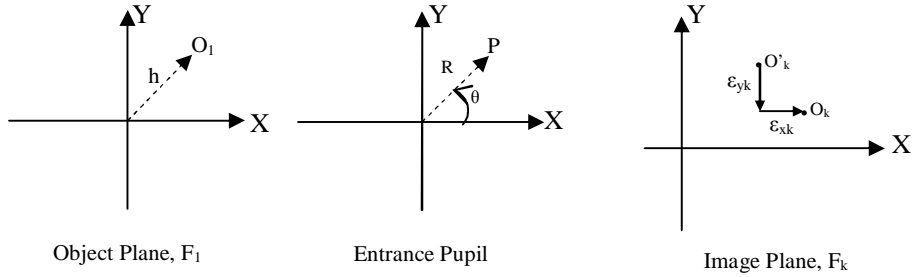

**Fig. 9** Coordinates in different planes used in the expansion of aberration terms (Description in the text)

$\varepsilon_k = \varepsilon_k^{(3)} + \varepsilon_k^{(5)} + \varepsilon_k^{(7)} + ...$, where $\varepsilon_k^{(r)}$ is the polynomial of degree $r$.

In terms of the coordinates (h, R, θ), the third order polynomial can be written as

$$\varepsilon_{yk}^{(3)} = \sigma_1 R^3 \cos\theta + \sigma_2(2 + 2\cos 2\theta)R^2 h + (3\sigma_3 + \sigma_4)Rh^2 \cos\theta + \sigma_5 h^3$$

$$\varepsilon_{xk}^{(3)} = \sigma_1 R^3 \sin\theta + \sigma_2 R^2 h \sin 2\theta + (\sigma_3 + \sigma_4)Rh^2 \sin\theta$$

where $(\sigma_1, \sigma_2, \sigma_3, \sigma_4, \sigma_5)$ are the third-order coefficients of the system.

Similarly $\varepsilon_k^{(5)}, \varepsilon_k^{(7)}$ can be expanded in (h, R, θ) with coefficients $(\mu_1, \mu_2, \mu_3, ...., \mu_{12})$ and $(\tau_1, \tau_2, \tau_3, .....\tau_{20})$ respectively (Cruickshank et al. 1960).

*Calculating Aberration Coefficients*
Buchdahl has shown that the coefficients of any order can be calculated by paraxial ray trace of two rays, a ray- marginal ray of on-axis object point and b ray- chief ray from the maximum radial field point. He also showed that successive order coefficients can be derived from the previous order coefficients in an iterative method (Buchdahl 1954, 1956, 1958). The same methodology is used to calculate the coefficients given the camera optics description and the rotationally symmetric camera optics.

*HESP Camera Implementation*

The HESP camera is a five element system, with first four rotationally symmetric elements and the final field flattener with a cylindrical surface. The first four elements are modelled as a single unit described in the above section (from now on this will be referred to as camera1). The output from the camera1 is traced through the field flattener to the detector plane. In order to get the complete ray coordinates from camera1 a ray's direction tangents are also to be determined. But the Buchdahl coefficients provide transverse aberrations alone.

Camera1 is modelled in two configurations as shown in Fig. 10. Configuration in Fig. 10(a), the object plane is at the original entrance pupil of the camera and the pupil plane at the first surface of the camera optics. Configuration in Fig. 10(b), the object plane is at infinity. The ABCD matrices and the aberration coefficients were determined for the two configurations of the camera1. A particular ray's coordinates on two different planes can be calculated and used, in turn, to calculate the tangents of the output ray of camera1.

As the objective of the model is to calculate the chief ray position on the detector plane, we deal with aperture sizes that cover chief rays of various wavelengths in the spectrograph. This also reduces the number of aberration terms required.

*Chromatic Corrections*

The wavelength dependency of refractive index is manifested in the wavelength dependency of the different coefficients. This dependency is captured in a relation in terms of wavelength.

Plots in Fig. 11 show the wavelength Vs. ABCD coefficients for camera1. The functional form,

$$\Lambda(\lambda) = \Lambda(\lambda_m) + (a\lambda + \frac{b}{\lambda^2} + \frac{c}{\lambda} + d)$$

where, $\Lambda$ can be any of A, B, C or D

$\lambda$ is wavelength,

$\lambda_m$ is a particular selected wavelength (in general mid wavelength of the band)

*a, b, c and d* are coefficients of the polynomial to fit

fits the relation between the coefficients and wavelengths to a coefficient of determination, R-square of 0.9999. This leaves just four parameters, *(A ($\lambda_m$), B ($\lambda_m$), C ($\lambda_m$) and D ($\lambda_m$))*, to include in the open parameters set for final optimization of the model to match the instrument. Similarly various aberration coefficients can be fit with the same functional form. But there will be no adjustable parameters considered for these relations.

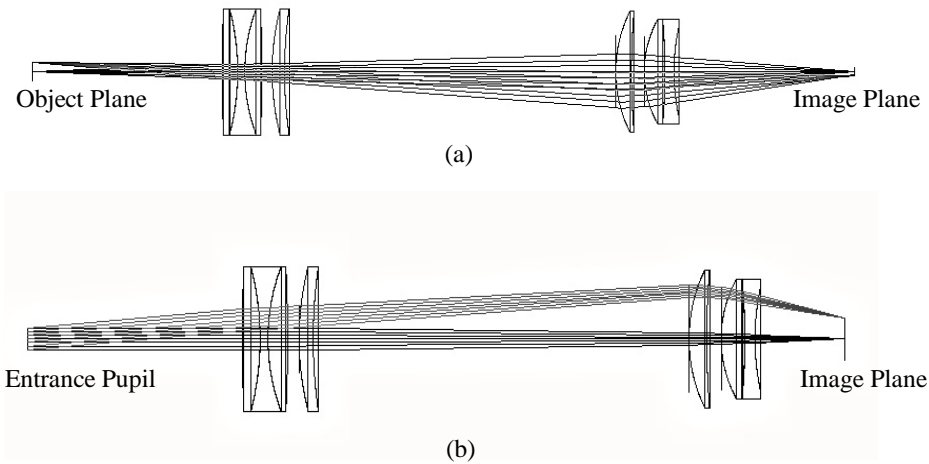

**Fig. 10** **(a)** Camera configuration with object plane at the camera system's entrance pupil in the spectrograph **(b)** Camera configuration with object plane at infinity

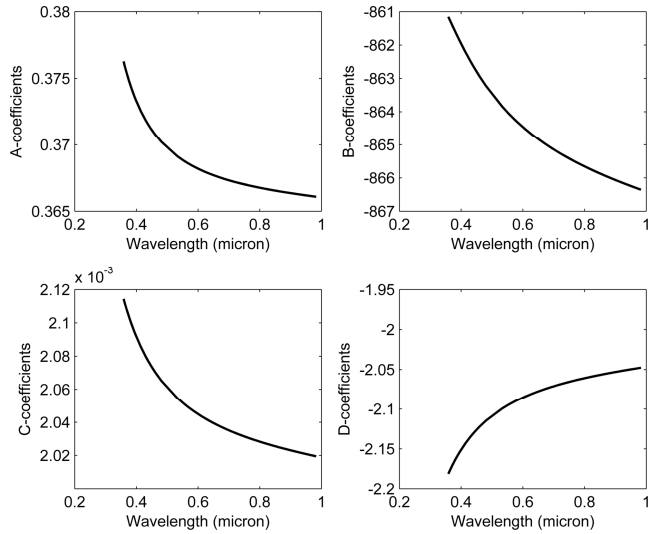

**Fig. 11** ABCD coefficients Vs. Wavelength for HESP camera1

The attributes to the camera class are the ABCD matrices and the aberration coefficients and the field flattener surface radii and its thickness. The class construct function includes the calculation of the ABCD and aberration coefficients when required. The transformation function's input are the ray coordinates on the first surface of the camera and the output is the ray coordinates on the last surface of the filed flattener.

6. Detector: Detector is defined by its pixel size and its dimensions. The function relevant converts the position on the image plane from distance units (millimetre) to pixels.

**2.3 Instrument Description**
The instrument is described by the components in proper order and the tilts and decentres associated with them. The chief ray trace starts with the ray description at the first component, the slit. An excel sheet describing the components and the distances between them will be read and the ray will be traced from one component to the next. The ray coordinates are transformed for the tilts associated by multiplying with the coordinate transform matrix (Ballester et al. 1997)

$$R = R_{\mu/x} R_{\nu/y} R_{\tau/z}$$

*where*

$$R_{\mu/x} = \begin{pmatrix} 0 & 0 & 1 \\ \cos\mu & \sin\mu & 0 \\ -\sin\mu & \cos\mu & 0 \end{pmatrix}$$

The refractive indices of various glasses for various wavelengths were calculated using the Selliemer coefficients and adjusted according to the environmental factors.

## 4. Comparison with Commercial Ray Trace Software

The model's performance is evaluated by comparing the output with the standard commercial ray trace software, Zemax. For a list of wavelengths, model and Zemax traces were done. The plots in Fig. 12 show the differences in x and y coordinates between Zemax and model for various orders of aberration terms included step by step. As expected, the contribution from the third order terms dominate. Using the terms till the seventh order, the difference is reduced to order of $10^{-5}$ mm.

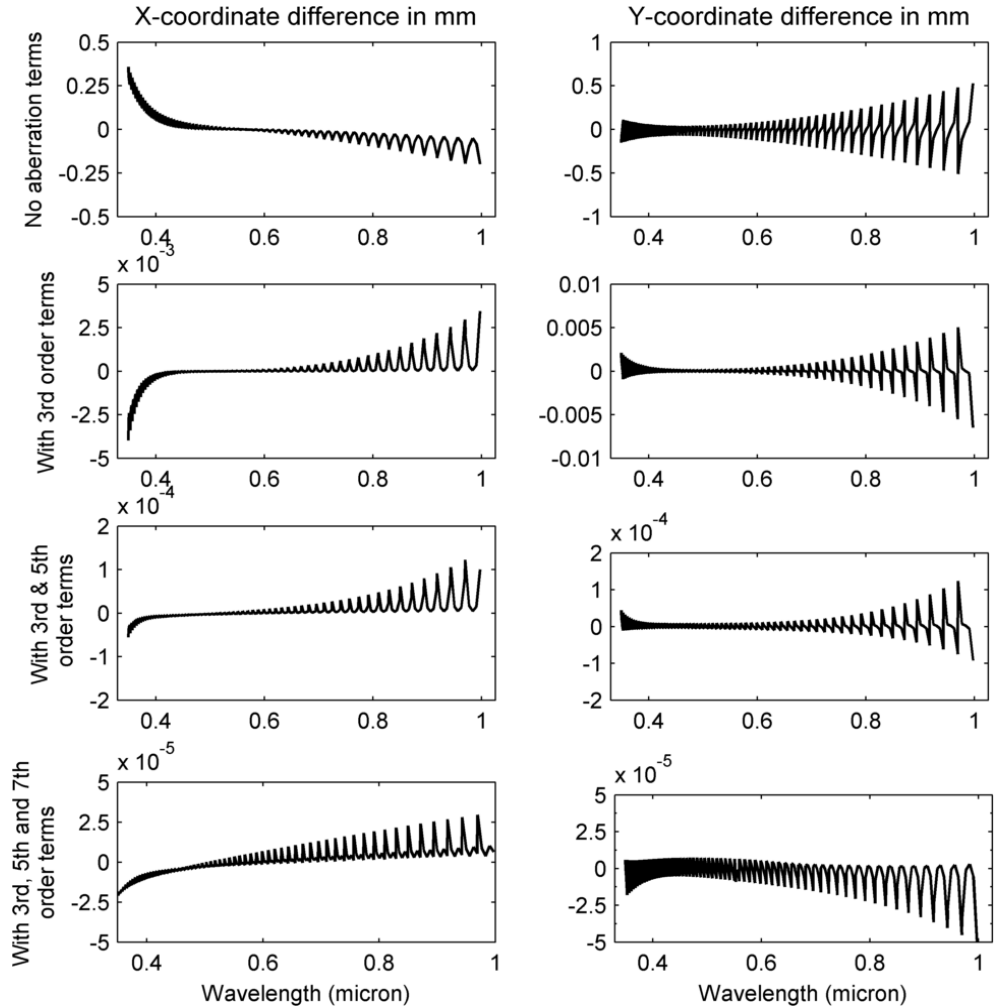

**Fig. 12** Plots showing the difference between Zemax trace and model for various aberration terms included. First column is the difference in x-coordinate, second column is the difference in y-coordinates. Each row of plots is difference in coordinates when various orders of aberration terms added as indicated in the far left of each row. X-axis of the plots is wavelength in microns, y-axis is the difference in the coordinates in mm

## 5. Matching Model with Instrument

To match the physical model with the built instrument, an optimization routine was introduced. Fig. 13 shows the flowchart of the process. Physical model being the core kernel of the routine, the ThAr exposures from the instrument are used along with the basic physical data like the environment variables, glass indices of the instrument. Some of the parameters of the model are left open, which include- slit orientation and position, collimator slit distance, Collimator tilt angles, Echelle tilt angles, slit mirror distance from collimator, fold mirror tilt angles, prism angles, camera ABCD parameters, and field flattener-CCD tilt angles, camera1 to field flattener distance, field flattener to CCD distance. The centroids of few selected features from the ThAr are used to optimize the open parameters to match with the model predictions. Given the possibility of multiple local minima, simulated annealing technique is used for optimization. First phase of model optimization has selective parameters open to study the effect of various terms on the model.

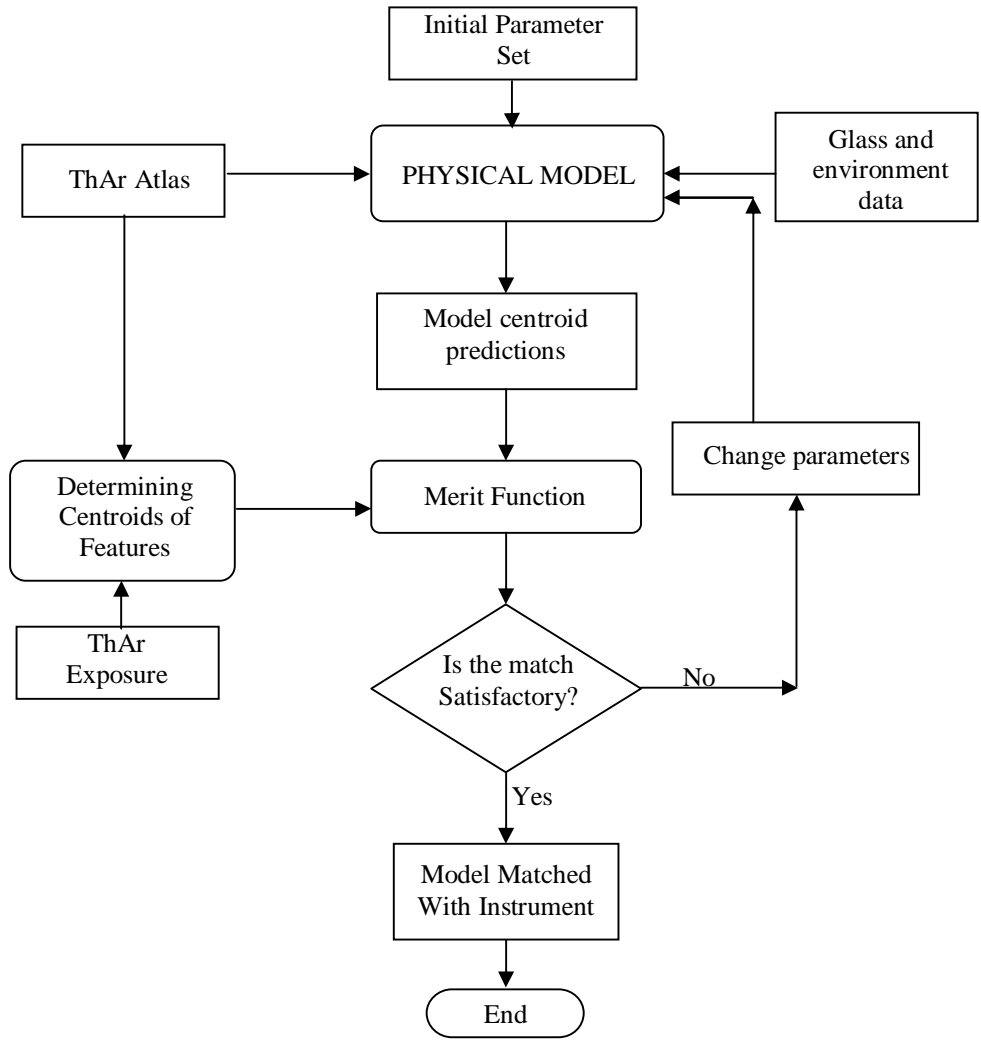

**Fig. 13** Flow chart of model optimization to match with the built instrument (Bristow et al. 2008)

**Choice of Feature Wavelengths**

It is important to know the accurate centroids and wavelengths of the features used. Though the ThAr atlases provide many lines, given the 30000 and 60000 resolutions of HESP, most of lines cannot be used due to blending which may cause inaccurate centroiding. A set of Thorium emission wavelengths, from the atlas by Redman et al. 2014, that are well isolated and strong enough to give accurate centroids is selected. A subimage of three times the resolution width is selected around the feature and 1D Gaussian function along with slope terms to include the effect of continuum is fit to the extracted 1D data in x and y directions to determine the centroid.

## 6. Preliminary Results

After an initial alignment of the HESP instrument, a few ThAr spectra were acquired and the model was tested. We present here the first results. Fig. 14 shows the match between model and the instrument for the set of wavelengths used for optimization. 92 features were selected for the optimization. Fig. 15 shows the scatter of residuals between the model's prediction and the centroids from the spectrum for a different set of 200 wavelengths to check for the accuracy of the model. The test wavelengths cover the entire spectrum and also same wavelength features in different orders.

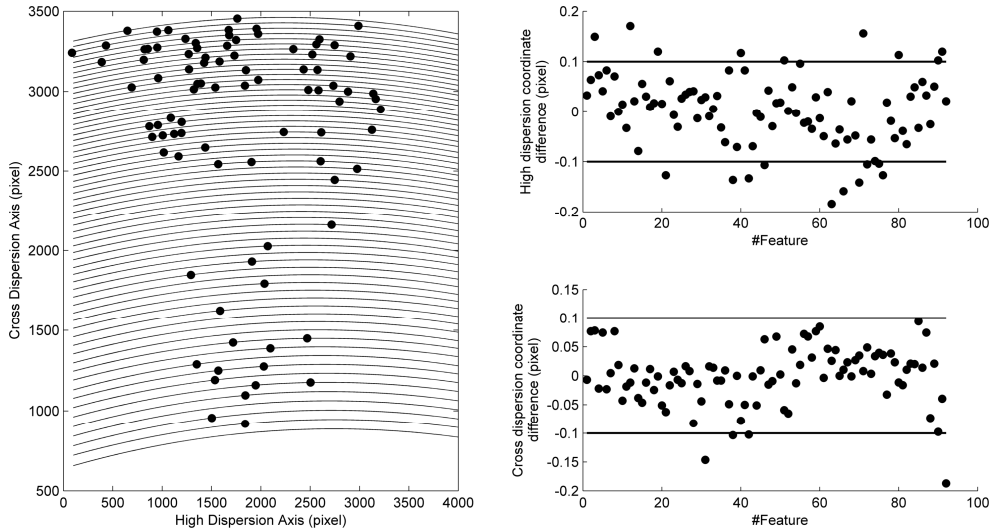

**Fig. 14** Scatter on the left shows the selected features' positions on CCD across different orders of ThAr spectrum from the instrument . The plots in the upper right panel and the lower right panel are the differences in high dispersion and cross dispersion coordinates for all the features respectively used for optimization

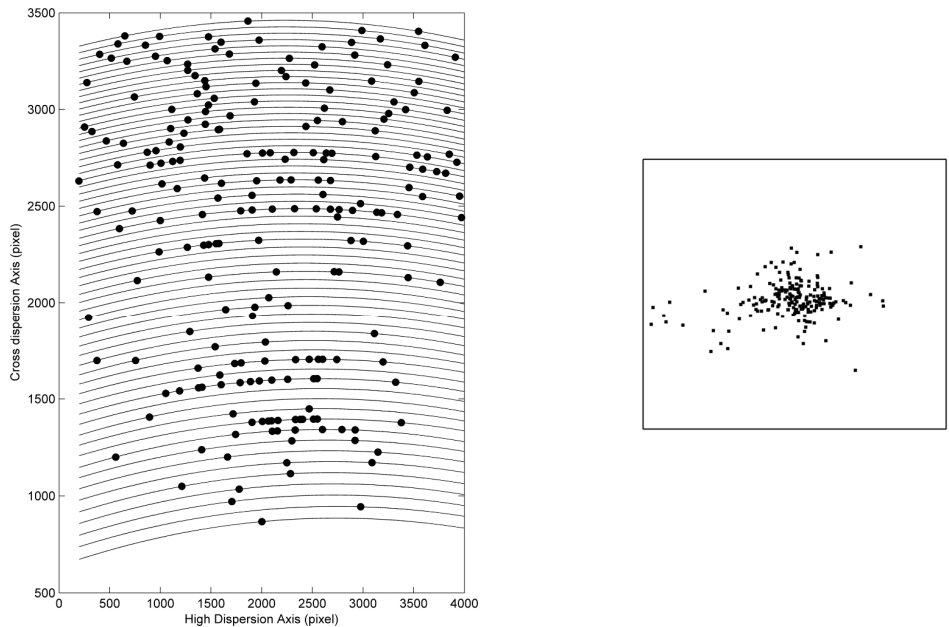

**Fig. 15** Left: Selected test features' positions across different orders. Right: Scatter of residuals for test features. The box represents one pixel

Subpixel accuracy is achieved with optimization of various parameters. The scatter of the residuals is mostly concentrated at the centre. The points that are far from the centre are mostly due to issues with centroiding of the features- either due to multiple features in the subimage that are unaccounted for or a very low signal to noise ratio of the feature. The effect of unaccounted multiple features in the subimage can be seen as the greater spread in the high dispersion direction (horizontal axis in Fig. 15) than the cross dispersion direction.

**Effect of Photon noise**
The measured centroids have uncertainties associated dependent on photon noise. Fig. 16 shows the histogram of the peak intensities of the ThAr lines when the strongest lines reach saturation on the CCD for HESP. 95% of the lines have peak intensities less than 2000 counts. For a well isolated line with no blends, Fig. 16 shows the uncertainties in the centroid for various signal levels, for a 4 pixel per resolution element sampling. Given the 350nm to 1000nm coverage on a single detector, the strong lines of Thorium in the red limit the photons collected in the mid and blue regions of the spectrum.

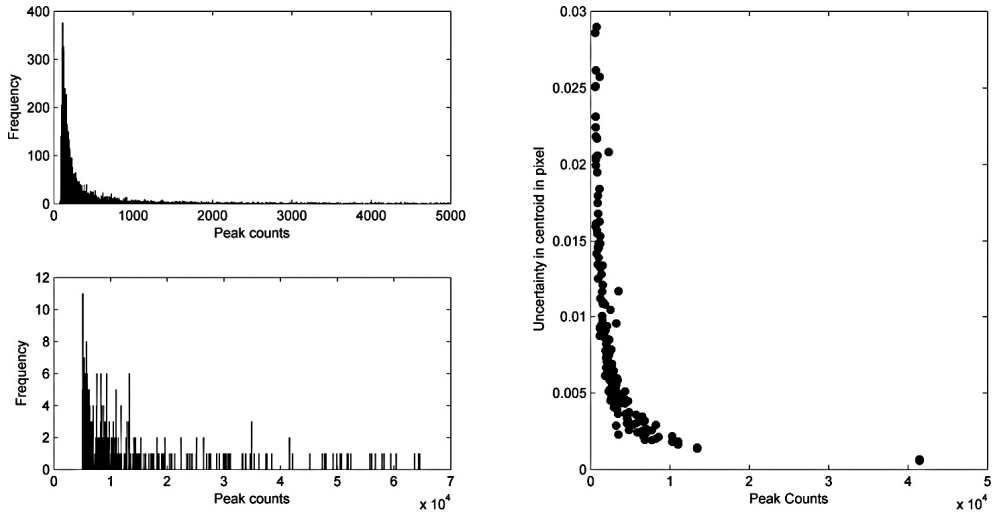

**Fig. 16** Column in the left shows histograms of peak counts of lines over the entire spectrum of HESP. Plot in the right column shows the uncertainty in the centroid through 2D Gaussian fit to Thorium lines vs. the peak intensity of the line

Bootstrap Monte Carlo simulations were run on the preliminary data from the instrument to check the model's performance to photon noise. Centroids were randomly subjected to the photon noise errors and the model was optimized every time. Standard deviation of model predictions of each feature was calculated. Fig. 17 shows the standard deviation of the model predictions and the standard deviation of centroids of the features subjected to photon noise.

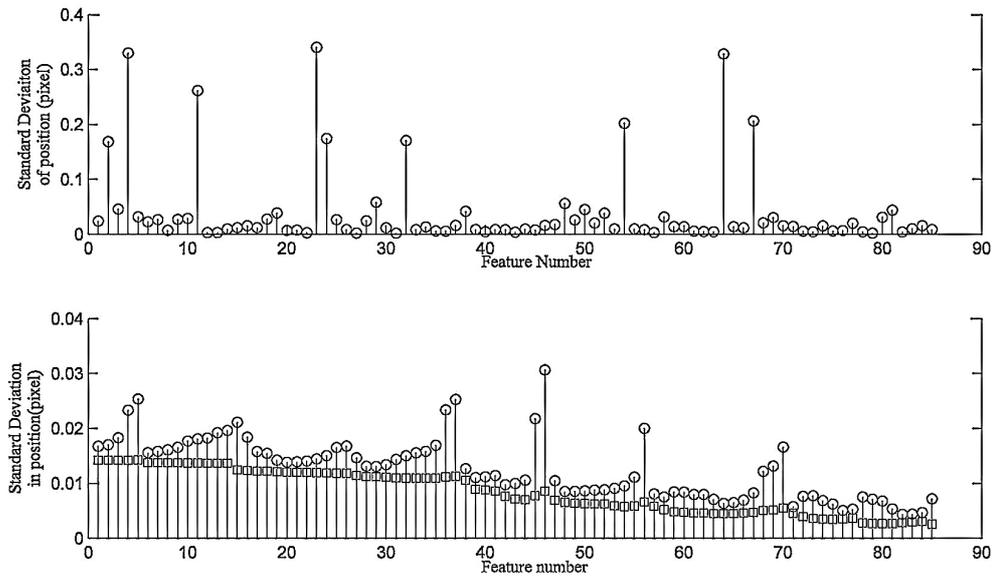

**Fig. 17** Top panel shows the centroid uncertainties (one sigma)of some of the test features. Bottom panel shows the uncertainties of the model predictions for two sets of bound conditions used (circles and squares). (Standard deviations in pixel units)

Almost constant errors for model predictions can be observed in the model predictions. The uncertainties occur due to the optimization routine used and the degenerate solutions possible for the model. Also the bounds on the model parameters and the signal to noise ratio of the features used for optimization contribute to the magnitude of the uncertainties. In Fig. 17 bottom panel, circles and squares represent the uncertainties for two sets of boundary conditions used on the open parameters for optimization, where the squares' parameters are more constrained than the circles'.

The test frames include the complete spectrum from 350nm to 1000nm. The strong Thorium lines in the red region limit the photons collected in the features. The photon count can be increased in the features when a subsection of the spectrum is of interest. Fig. 18 shows the effect of better signal strength of the image features. The circles and squares represent the same as Fig. 17 . The uncertainties have reduced considerably in the simulations done. A gradual change of uncertainties with order and a variation within an order itself can be observed. These are the direct results of the instrument behaviour for changes in parameters. This sort of knowledge about the variations can be very useful in ThAr simultaneous observations for drift corrections, where usually a constant drift across the spectrum is assumed for corrections.

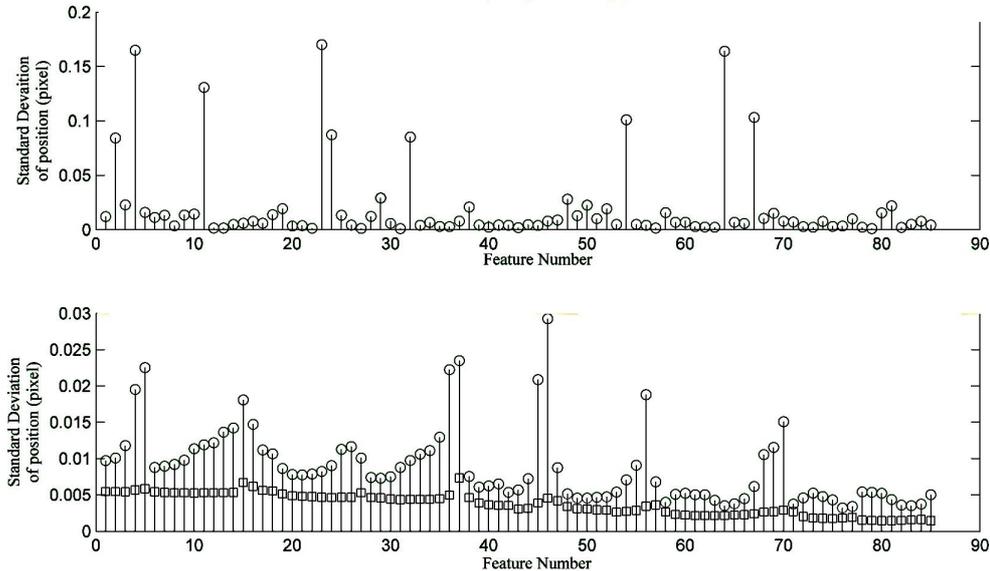

**Fig. 18** Top panel shows the centroid uncertainties (one sigma) of some of the test features. Bottom panel shows the uncertainties of the model predictions for two sets of bound conditions used (circles and squares). (Standard deviations in pixel units)

As more lab tests on the instrument are done, the parameters can be well constrained and if not all, some of the degenerate cases can be eliminated resulting in better performance of the model. Further work on this will be presented in the follow up paper in preparation.

## 7. Summary

A physical model was developed for the Hanle Echelle Spectrograph and tested with the preliminary alignment data from the lab. We have shown that such a model can be built to a high degree of accuracy. Using very few lines from the ThAr spectrum, any wavelength position can be predicted in the spectrum. This will predict any number of lines across the spectrum to any required density. As seen in the results for the photon noise effect on the predictions, almost consistent uncertainties can be achieved for the predictions unlike the varying centroid uncertainties of the CCD image features. More tests during alignment will provide more information about the instrument to add to the model and more accurate data that will lead to better predictions.

The technique is such that it can be implemented without the use of any commercial softwares and on any regular system available at the observatories. This approach facilitates a better understanding of the built instrument from design and supports calibrations. Another application in scope is its use in ThAr simultaneous calibrations. The drift corrections applied in simultaneous reference observations assumes that the star and the calibration light follow the same path in the instrument and the drift is constant over the entire spectrum, this is not an accurate assumption for very high precision spectroscopy. A physical model will be able to simulate position dependent shifts in the instrument and hence a more accurate drift corrections. The follow up paper will present in detail the performance of the model in conjunction with the instrument. Once an easy to use frame work is built around the model, the code will be made available to anyone interested. Meanwhile we are open for collaborations with any teams interested to develop a similar model for their spectrographs.